# Have LLM-associated terms increased in article full texts in all fields?

Mike Thelwall, School of Information, Journalism and Communication, University of Sheffield, UK. https://orcid.org/0000-0001-6065-205X;

Kayvan Kousha, Statistical Cybermetrics and Research Evaluation Group, Business School, University of Wolverhampton, UK. https://orcid.org/0000-0003-4827-971X

The use of Large Language Models (LLMs) like ChatGPT and DeepSeek for translation and language polishing is a welcome development, reducing the longstanding publishing barrier to non-English speakers. Assessing the uptake of this facility is useful to give insights into changing nature of scientific writing. Although the prevalence of LLM-associated terms has been tracked across science in abstracts and for full text biomedical research, their science-wide prevalence in full texts is unknown. In response, this article investigates an expanded set of 80 potentially LLM-associated terms during 2021-2025 in a science-wide full text collection from the publisher MDPI (1.25 million articles), partly focusing on the 73 journals that published at least 500 articles in 2021. The results demonstrate the increasing prevalence of LLM-associated terms science-wide in full texts to 2024, with some terms declining from 2024 to 2025 and others continuing to increase. LLMs seem to avoid some terms (e.g., thus, moreover) and a few terms have stronger associations with abstracts than full texts (e.g., enhanced) or the opposite (e.g., leveraged). The term family "underscore" had the biggest increase: up to 29-fold. There are substantial differences between journals in the apparent use of LLMs for writing, from lower uptake in the life sciences to higher uptake in social sciences, electronic engineering and environmental science. Fields in which there is currently low uptake may need improved or specialist support, such as for reliably translating complex formulae, before the full benefits of automatic translation can be realised.
**Keywords**: Large language models; LLMs; Academic writing; AI language; LLM-associated terms; full-text analysis

## Introduction

Science publishing has always been linguistically exclusionary of those not fluent in the dominant languages, from Arabic to Roman and Greek in the European renaissance to English today (e.g., Editorial, 2023). Whilst minority publishing in other languages has continued in parallel and even flourished (e.g., French for sociology; Chinese for 红学; much humanities research in national languages) it does not seem to be highly regarded in some areas of science (Di Bitetti, & Ferreras, 2017), even though it can contain essential information (Angulo et al., 2021). Now, however, Large Language Models (LLMs) like DeepSeek and ChatGPT have international use in academia (Mishra et al., 2024; Mohammadi et al., 2026; Van Noorden & Perkel, 2023) and seem to offer cheap and effective translation capabilities. This can reduce the language barrier by translating the scholarly literature into the main languages of non-English speakers for literature reviews and helping authors to translate their work into English or polish their initial English drafts (Khalifa & Albadawy, 2024; Lechien et al., 2024; Wu et al., 2026). It is not known whether this is equally effective in all fields, however. If not, then alternative actions may be needed.

LLM translations do not closely mimic academic style in some respects; They can introduce terms that are either not fully appropriate or that are unusual in scholarly writing. Identifying and tracking these terms can give indirect evidence of the use of LLMs in academic writing over time (e.g., Comas-Forgas et al., 2025; Gray, 2024; Juzek & Ward, 2025; Kobak et al., 2024; Liang et al., 2024; Uribe & Maldupa, 2024). Linguistic complexity can also be used to track LLM language changes (Alsudais, 2025; Bao et al., 2025). The most recent and largest scale study tracked 12 term families in titles/abstracts from six scholarly databases from 2015 to 2024, finding the largest increases for *delve* (15-fold increase), *underscore* (10-fold), and *intricate* (7-fold). Disciplinary differences were found, with larger increases in STEM fields than in the social sciences, arts and humanities. The term *underscore* also greatly increased in prevalence by 2024 in PubMed Central (PMC) biomedical full text papers (Kousha & Thelwall, 2026). Despite this, the extent of use of LLMs in the full text of journal articles science wide is unknown.

This article assesses the prevalence of LLM-associated terms in the full text of academic papers science-wide to identify the overall presence and any disciplinary differences. Since abstracts are easier to obtain for analysis than full texts, this article also assesses whether LLM term prevalence differs between the two: a negative answer would allow abstracts to be used as a proxy for full texts in future studies. The following research questions address these issues.

- RQ1: Have LLM-associated terms increased in prevalence in full texts science-wide?
- RQ2: Does the prevalence of LLM-associated terms in full texts mirror their presence in abstracts?
- RQ3: Are there field differences in LLM-associated term prevalence?

## Methods

The research design was to obtain a large science-wide collection of full text journal articles classified by discipline, count the LLM-associated terms in them, and then compare results between disciplines. LLM-usage declaration statements in articles were not checked as probably most authors do not use them (Wu et al., 2026).

### *Data: Articles*

There are no public science-wide English-language collections of full text journal articles suitable for text processing. Although article abstracts are routinely and systematically available either through CrossRef or through citation databases, such as OpenAlex, Scopus and Dimensions, the same is not true for full texts. Instead, they can be paywalled or, if open access, are available in diverse formats (e.g., PDF, HTML). Moreover, effective processing of PDFs and HTML from multiple publishers is difficult because it involves removing publisher-specific information, page layout features and watermarks. For biomedical research there is a useful source of article full texts in XML format within PubMed Central (Roberts, 2001) but there is no equivalent science-wide source. For example, preprint repositories tend to have a subject focus (e.g., arXiv), institutional repositories contain documents in varied formats, and large-scale web crawling databases also use many formats, can have incomplete metadata, and are not systematic.

For this study, journal articles from the publisher MDPI were used as the raw data because they seem to encompass all areas of science and are available in XML format and so can be easily and accurately processed. The XML files were downloaded in December 2025 with permission from the publisher and processed to extract each article's title, abstract and main text, excluding references and appendices. This source has apparently error-free metadata (journal, publication year) as well as error-free full text because it is the official source rather than a processed copy. We did not test for errors systematically have conducted hundreds of ad-hoc checks of individual articles, finding no problems.

### *Data: LLM-associated terms*

A list of 80 LLM-associated terms was generated to identify in the journal articles dataset. This was compiled from previous studies (Kousha & Thelwall, 2026), with word stems expanded into complete words. In addition, ChatGPT 5.2 and Gemini 3 were asked for LLM-associated terms in case they could identify any that had been previously overlooked. They suggested "complex tapestry", which was retained even though it is a multi-word phrase. The other LLM-suggested terms that had not been previously included were: moreover, furthermore, thus, and importantly, which were claimed to be "Overused transition and structuring phrases". Every term was searched for in each MDPI journal article, reporting the binary decision of whether it was present or not.

### *Analysis*

Articles were analysed by journal rather than by any other field categorisation. This is not ideal because MDPI journals tend to be multidisciplinary to some extent. Nevertheless, this seems more transparent than attempting to classify the articles individually, because individual article-level classification errors would not be evident to the reader.

Based on all the above, the raw data was the proportion of articles in each journal containing each term, obtained separately for each year. The extent to which a term is LLM-associated was calculated in several ways, all based on the same assumption: that any increase in the proportion of articles containing a given term was due to LLM use by authors. The year 2021 was used as the reference year because the first major GPT, ChatGPT was released to the public at the end of 2022. The most recent (almost) complete year, 2025, was chosen as the end year. Of course, there are many reasons why language use can change, including the emergence of new hot topics, so increases may not be due to LLMs.

Using $p_n$ to represent the proportion of articles containing a given term in year $n$, the first calculation was the increase in the proportion of articles in the journal using the term: $p_{2025}-p_{2021}$. This reflects the prevalence in the journal of the term from LLMs, on the assumption that LLMs have caused the increase. Second, $p_{2025}/p_{2021}$ is the ratio of the prevalence of a term in 2025 to its pre-LLM prevalence. Finally, $p(LLM|term)=(p_{2025}-p_{2021})/p_{2025}$ is the probability that a term in a 2025 article was written by a LLM, if the increase is solely due to LLMs.

# Results

## RQ1: LLM-associated term prevalence in science-wide full texts

Most of the 80 terms tested had a substantial increase in prevalence from 2021 to 2025, whether measured by percentage increase or increase ratio (Table 1). For absolute prevalence throughout all full texts, *enhance* is dominant, occurring in almost two thirds (65.3%) of full texts in 2025, an increase of 37.2% from 2021, and 2.3 times more common in 2025 than 2021. The *underscore* word family had the largest increase ratio, with *underscores* emerging from academic obscurity (0.8%) to common use (22.8%), a 28.9-fold increase.

Some of the words claimed by ChatGPT to be LLM-associated were poor choices, with decreasing prevalence. These included terms that are archaic outside of an academic setting (*thus*, *moreover*, *furthermore*) as well as *importantly* and *unveil*. There was little change overall to the unveil family of words. The term *importantly*, it may have been replaced by the similar (more archaic) term *notably*.

Table 1. The prevalence of the 80 proposed LLM-associated terms in MDPI journal **full texts** of articles during 2021-2025. p(LLM|term) is the probability of LLM influence being the cause of an individual term, assuming no other influences on language change (n=237,215 in 2021 to n=264,835 in 2025) (see also Table A1 for comparison).

| Proportion | 2021 | 2022 | 2023 | 2024 | 2025 | Increase | Increase ratio | p(LLM|term) |
|---|---|---|---|---|---|---|---|---|
| enhance | 28.1% | 29.2% | 37.1% | 52.1% | 65.3% | 37.2% | 2.3 | 0.57 |
| crucial | 29.8% | 30.0% | 41.7% | 59.5% | 60.6% | 30.8% | 2.0 | 0.51 |
| comprehensive | 21.3% | 22.9% | 31.3% | 46.0% | 55.9% | 34.5% | 2.6 | 0.62 |
| thus | 58.6% | 57.1% | 57.2% | 56.5% | 52.8% | -5.8% | 0.9 | -0.11 |
| enhanced | 29.4% | 29.6% | 33.3% | 41.6% | 52.5% | 23.1% | 1.8 | 0.44 |
| enhancing | 14.6% | 15.4% | 22.4% | 39.7% | 51.4% | 36.8% | 3.5 | 0.72 |
| robust | 16.7% | 16.0% | 19.3% | 28.6% | 41.8% | 25.1% | 2.5 | 0.60 |
| enhances | 10.5% | 10.7% | 14.9% | 27.3% | 40.9% | 30.5% | 3.9 | 0.74 |
| facilitate | 19.9% | 19.0% | 22.8% | 30.1% | 35.6% | 15.7% | 1.8 | 0.44 |
| facilitating | 7.3% | 6.7% | 11.0% | 21.8% | 26.4% | 19.1% | 3.6 | 0.72 |
| facilitates | 8.4% | 7.8% | 10.8% | 18.0% | 24.0% | 15.6% | 2.8 | 0.65 |
| underscores | 0.8% | 0.7% | 3.8% | 15.6% | 22.8% | 22.0% | 28.9 | 0.97 |
| underscore | 0.8% | 0.7% | 2.9% | 11.7% | 21.0% | 20.2% | 26.8 | 0.96 |
| emphasize | 6.3% | 5.9% | 8.1% | 12.7% | 21.0% | 14.7% | 3.3 | 0.70 |
| innovative | 9.6% | 9.4% | 12.6% | 19.1% | 20.7% | 11.0% | 2.1 | 0.53 |
| emphasizing | 3.2% | 3.1% | 5.5% | 13.7% | 19.8% | 16.6% | 6.2 | 0.84 |
| notably | 5.7% | 5.4% | 8.0% | 15.9% | 19.3% | 13.6% | 3.4 | 0.70 |
| pivotal | 5.2% | 4.5% | 8.3% | 17.9% | 18.0% | 12.9% | 3.5 | 0.71 |
| underscoring | 0.4% | 0.4% | 1.8% | 8.3% | 17.2% | 16.8% | 41.0 | 0.98 |
| emphasizes | 4.0% | 4.0% | 6.1% | 11.4% | 16.3% | 12.3% | 4.1 | 0.76 |
| leveraging | 1.2% | 1.2% | 3.9% | 10.6% | 15.7% | 14.5% | 13.0 | 0.92 |
| emphasized | 7.7% | 7.4% | 8.4% | 11.0% | 15.2% | 7.6% | 2.0 | 0.50 |
| facilitated | 6.5% | 6.0% | 8.3% | 13.2% | 13.9% | 7.4% | 2.1 | 0.53 |
| interplay | 3.9% | 3.2% | 5.2% | 9.8% | 13.2% | 9.3% | 3.4 | 0.70 |
| intricate | 1.7% | 1.6% | 6.2% | 13.5% | 10.7% | 9.0% | 6.3 | 0.84 |

| word | | | | | | | | |
|---|---|---|---|---|---|---|---|---|
| leverage | 2.0% | 2.0% | 3.5% | 7.0% | 10.3% | 8.3% | 5.1 | 0.80 |
| paradigm | 5.1% | 4.8% | 5.2% | 5.9% | 10.1% | 5.1% | 2.0 | 0.50 |
| fostering | 1.6% | 1.4% | 3.2% | 7.6% | 9.8% | 8.2% | 6.2 | 0.84 |
| heightened | 1.4% | 1.3% | 3.8% | 8.9% | 9.7% | 8.2% | 6.8 | 0.85 |
| leverages | 0.6% | 0.6% | 2.0% | 5.4% | 8.7% | 8.1% | 14.7 | 0.93 |
| multifaceted | 1.8% | 1.7% | 3.4% | 7.6% | 8.6% | 6.8% | 4.7 | 0.79 |
| profound | 4.3% | 4.0% | 5.0% | 7.4% | 8.5% | 4.2% | 2.0 | 0.49 |
| nuanced | 0.9% | 0.8% | 2.1% | 6.7% | 8.2% | 7.3% | 9.4 | 0.89 |
| foster | 2.8% | 2.4% | 3.4% | 5.6% | 8.0% | 5.2% | 2.8 | 0.65 |
| synergy | 2.4% | 2.4% | 2.7% | 3.6% | 6.7% | 4.4% | 2.8 | 0.65 |
| importantly | 7.3% | 6.8% | 6.2% | 5.4% | 6.0% | -1.4% | 0.8 | -0.23 |
| transformative | 0.7% | 0.6% | 1.3% | 3.0% | 5.6% | 4.9% | 7.7 | 0.87 |
| leveraged | 1.1% | 1.1% | 1.7% | 3.1% | 4.6% | 3.6% | 4.4 | 0.77 |
| fosters | 0.7% | 0.7% | 1.2% | 3.0% | 4.6% | 3.8% | 6.2 | 0.84 |
| paradigms | 1.4% | 1.3% | 1.5% | 1.9% | 4.5% | 3.1% | 3.2 | 0.69 |
| navigate | 1.2% | 1.1% | 1.9% | 3.6% | 4.1% | 2.9% | 3.5 | 0.71 |
| showcasing | 0.2% | 0.2% | 1.7% | 5.4% | 3.8% | 3.6% | 16.4 | 0.94 |
| underscored | 0.4% | 0.4% | 1.3% | 3.8% | 3.8% | 3.3% | 8.4 | 0.88 |
| meticulously | 0.3% | 0.3% | 1.8% | 5.0% | 3.1% | 2.9% | 11.3 | 0.91 |
| navigating | 0.6% | 0.6% | 1.0% | 2.2% | 2.7% | 2.1% | 4.4 | 0.77 |
| nuances | 0.6% | 0.5% | 1.1% | 2.4% | 2.5% | 1.9% | 4.3 | 0.77 |
| robustly | 1.0% | 0.9% | 1.0% | 1.2% | 2.3% | 1.3% | 2.3 | 0.56 |
| meticulous | 0.4% | 0.4% | 1.5% | 3.5% | 2.3% | 1.9% | 5.3 | 0.81 |
| harness | 0.8% | 0.7% | 1.4% | 2.4% | 2.2% | 1.5% | 3.0 | 0.67 |
| delve | 0.5% | 0.5% | 2.3% | 4.9% | 2.1% | 1.7% | 4.6 | 0.78 |
| moreover | 3.9% | 3.8% | 3.4% | 2.7% | 2.0% | -1.9% | 0.5 | -0.97 |
| synergies | 0.9% | 0.7% | 0.8% | 1.1% | 1.8% | 0.9% | 2.0 | 0.50 |
| bolster | 0.3% | 0.3% | 0.9% | 2.1% | 1.6% | 1.2% | 4.8 | 0.79 |
| fostered | 0.8% | 0.7% | 0.8% | 1.0% | 1.6% | 0.7% | 1.9 | 0.47 |
| showcases | 0.3% | 0.3% | 1.5% | 2.9% | 1.6% | 1.3% | 5.6 | 0.82 |
| harnessing | 0.5% | 0.5% | 1.3% | 2.3% | 1.5% | 1.0% | 3.0 | 0.67 |
| showcase | 0.6% | 0.5% | 1.3% | 2.3% | 1.5% | 0.9% | 2.6 | 0.61 |
| furthermore | 3.0% | 2.6% | 2.5% | 2.0% | 1.5% | -1.5% | 0.5 | -1.01 |
| delves | 0.1% | 0.1% | 1.2% | 3.4% | 1.4% | 1.3% | 13.8 | 0.93 |
| groundbreaking | 0.3% | 0.2% | 0.7% | 1.4% | 1.2% | 0.9% | 4.6 | 0.78 |
| heighten | 0.2% | 0.2% | 0.3% | 0.6% | 1.1% | 0.9% | 4.9 | 0.80 |
| showcased | 0.3% | 0.4% | 1.2% | 2.1% | 1.0% | 0.7% | 3.2 | 0.69 |
| harnessed | 0.5% | 0.4% | 1.0% | 1.2% | 1.0% | 0.5% | 2.0 | 0.49 |
| nuance | 0.2% | 0.2% | 0.2% | 0.3% | 0.9% | 0.7% | 4.0 | 0.75 |
| crucially | 0.7% | 0.6% | 0.6% | 0.7% | 0.8% | 0.1% | 1.1 | 0.11 |
| bolstering | 0.1% | 0.1% | 0.6% | 1.5% | 0.8% | 0.7% | 7.7 | 0.87 |
| unveiled | 0.7% | 0.6% | 1.3% | 2.1% | 0.8% | 0.1% | 1.1 | 0.09 |
| delved | 0.1% | 0.1% | 1.0% | 2.2% | 0.7% | 0.6% | 5.7 | 0.82 |
| delving | 0.2% | 0.2% | 0.7% | 1.6% | 0.7% | 0.5% | 3.7 | 0.73 |
| heightens | 0.1% | 0.1% | 0.2% | 0.4% | 0.7% | 0.6% | 6.5 | 0.85 |
| commendable | 0.1% | 0.1% | 0.7% | 1.4% | 0.6% | 0.5% | 5.9 | 0.83 |
| heightening | 0.1% | 0.1% | 0.2% | 0.4% | 0.6% | 0.5% | 5.3 | 0.81 |
| unveil | 0.7% | 0.6% | 0.9% | 1.3% | 0.6% | -0.2% | 0.8 | -0.27 |
| paradigmatic | 0.3% | 0.2% | 0.2% | 0.3% | 0.4% | 0.1% | 1.4 | 0.30 |
| navigated | 0.2% | 0.2% | 0.3% | 0.4% | 0.4% | 0.2% | 2.1 | 0.53 |
| bolsters | 0.1% | 0.1% | 0.2% | 0.5% | 0.4% | 0.3% | 6.1 | 0.84 |
| unveiling | 0.3% | 0.2% | 0.4% | 0.7% | 0.3% | 0.1% | 1.2 | 0.19 |
| navigates | 0.1% | 0.1% | 0.2% | 0.4% | 0.3% | 0.2% | 2.9 | 0.66 |
| unveils | 0.2% | 0.1% | 0.4% | 0.7% | 0.3% | 0.1% | 1.8 | 0.46 |
| complex tapestry | 0.0% | 0.0% | 0.0% | 0.0% | 0.0% | 0.0% | 8.4 | 0.88 |
| [Decr. freq. terms] | | 61 | 5 | 4 | 30 | | | |

It is evident in the data that the prevalence of LLM-associated terms generally increased sharply from 2022 to 2023 and from 2023 to 2024, but many (36%: 30 out of 80 terms; see the last row of Table 1) became less common in 2025. For example, *intricate* increased from 1.6% in 2022 to 13.5% in 2024 but then decreased to 10.7%.

### RQ2: Term prevalence in abstracts and full texts

The prevalence of LLM-associated terms in abstracts was of course less than in full texts: about a tenth as common overall (Appendix: Table A1). The increase ratios tended to be slightly higher for full texts (median: 3.5 to 1 between 2021 and 2025) than for abstracts (median: 2.9 to 1 between 2021 and 2025) so LLMs may be more important for full texts than for abstracts and are certainly not restricted to generating summaries for abstracts.

For most terms the probability of being LLM-generated (under the assumptions given of only LLMs influencing language changes) is broadly similar for both full texts and abstracts, but there are some substantial differences (Figure 1). In particular, *enhance* family words are more strongly associated with LLMs in abstracts than in full texts, but the opposite is true for *robustly*, *leveraged,* and *showcase*.

In addition to differences in association strengths, there are also terms that have increased in prevalence in full texts but decreased in prevalence in abstracts: *harnessed*, *unveiling*, *unveiled*, and *crucially*. Overall, therefore, abstract presence is a poor guide to full text prevalence for individual LLM-associated terms, but more reliable in aggregate.

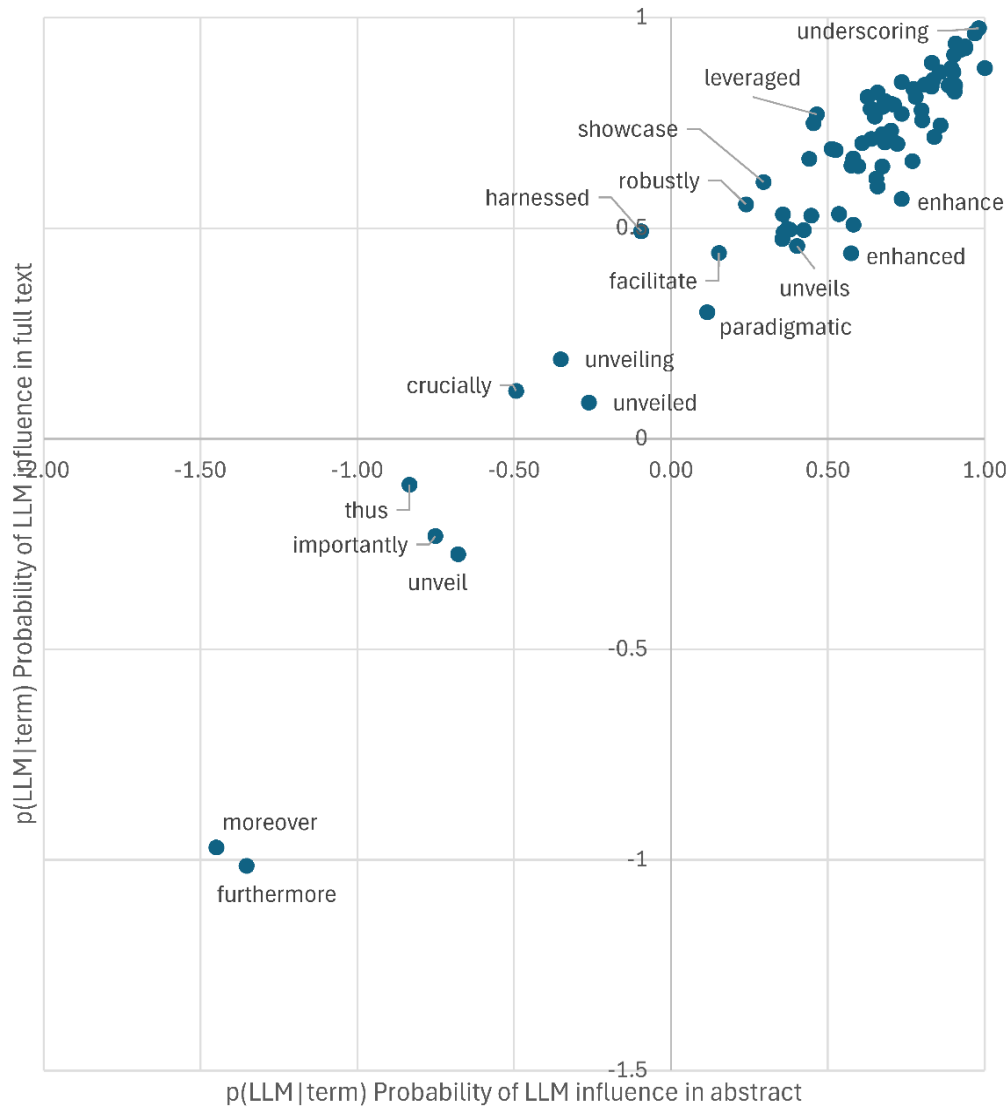

Figure 1. The probabilities of 80 terms being LLM-generated in MDPI journal full text articles against abstracts 2021-2025. The probabilities are based on the assumptions in the text. Negative values are not probabilities but calculated with the same formula used to calculate the probabilities.

## RQ2: Field/journal differences

There were substantial differences between journals in the incorporation of LLM-associated terms in full texts, at least from an average frequency increase perspective (Table 2). For example, whilst the 80 terms checked increased in average prevalence by 2% for *Universe*, they increased six times more,12% on average, in the journals *Information* and *Journal of Risk and Financial Management*. No journals experienced a decrease in LLM term prevalence between 2021 and 2025.

There are some broad disciplinary differences in uptake between journals. In particular, life science journals tend to have lower increases in LLM-associated terms and social sciences (*Information and Journal of Risk and Financial Management, ISPRS International Journal of Geo-Information; Education*), environmental science (*Sustainability; Environmental Science*) and electronic engineering (e.g., *Electronics,*

*Remote Sensing, Biosensors, Sensors*) tend to have higher increases in LLM-associated terms.

Table 2. Mean percentage increase in occurrence 2021-25 across all 80 terms in Table 1 by journal in article **full texts** for the 73 MDPI journals that had published at least 500 articles in 2021.

| Journal* | Mean term prevalence increase |
|---|---|
| Information, J. Risk & Financial Man. | 12% |
| Sustainability, ISPRS Int. J. Geo-Information | 11% |
| Land, Education, Electronics | 10% |
| **Remote Sensing** | 9% |
| Biosensors, Healthcare, Applied Sciences, Agriculture, Sensors, Symmetry, Buildings, Energies | 8% |
| J. Marine Science & Eng., Mathematics, Catalysts, Forests, **Int. J. Environ. Res. & Pub. Health**, Water, Processes, Pharmaceutics, Pharmaceuticals, Foods, Atmosphere, Agronomy, **Brain Sciences**, Diagnostics | 7% |
| Biomolecules, Nanomaterials, Antioxidants, J. Personalized Medicine, Medicina, Horticulturae, **Biology**, Marine Drugs, Polymers, Plants, Entropy, Micromachines, Nutrients, **Children**, Life, Microorganisms, Antibiotics, Membranes, Coatings | 6% |
| ***Vaccines***, Materials, Photonics, Geosciences, Int. J. Molecular Sciences, Cells, Biomedicines, J. Clinical Medicine, Genes, Metals, Molecules, Metabolites, Crystals, Religions, Minerals, J. Fungi, Animals, Toxins | 5% |
| Pathogens, Cancers, Viruses, Diversity, Insects | 4% |
|  | 3% |
| Universe | 2% |

*****Bold**: terms relatively common in full texts compared to abstracts; ***Bold italic***: terms rare in full texts compared to abstracts. These indicate a ranking change of at least 20 in Table 2 compared to Table 3.

There were also substantial differences between journals in the incorporation of LLM-associated terms in abstracts from an average prevalence increase perspective (Table 3). For individual journals, mean increases in prevalence differed substantially between abstracts and full texts. For example, for *Remote Sensing*, LLM-associated terms increased relatively more in full texts (9%) than abstracts (0.7%). Similarly, for *Biology*, LLM-associated terms increased relatively more in full texts (6%, ranked 8th highest) than in abstracts (0.3%, ranked 4th lowest). In contrast, for *Vaccines*, LLM-associated terms increased relatively more in abstracts (0.8%, ranked 27th) than in full texts (5%, ranked 50th).

Table 3. Mean percentage increase in occurrence 2021-25 across all the 80 terms in Table 1 by journal in article **abstracts** for the 73 MDPI journals that had published at least 500 articles in 2021.

| Journal* | Mean term prevalence increase |
|---|---|
| Information, Sustainability | 1.2% |
| Education, Electronics | 1.1% |
| J. Risk & Financial Man., Land, ISPRS Int. J. Geo-Information, Symmetry, Biosensors, Agriculture | 1.0% |
| Catalysts, Agronomy, Pharmaceutics, Sensors, Energies, Applied Sciences | 0.9% |
| J. Marine Science & Eng., Horticulturae, Processes, Healthcare, Plants, Nanomaterials, Mathematics, Buildings, Foods, Forests, **Vaccines**, Water | 0.8% |
| ***Remote Sensing***, Pharmaceuticals, Polymers, Atmosphere, Antioxidants, Diagnostics, Marine Drugs, Coatings, Biomolecules, J. Personalized Medicine, Membranes, Antibiotics, Micromachines, ***Int. J. Environ. Res. & Pub. Health***, Microorganisms, Entropy | 0.7% |
| Genes, Materials, Photonics, Metals, Geosciences, **Brain Sciences**, Life, Nutrients, Molecules, Medicina, J. Fungi | 0.6% |
| Metabolites, Int. J. Molecular Sciences, Minerals, J. Clinical Medicine, Crystals, Diversity, Religions, Biomedicines, Pathogens, Toxins, Viruses | 0.5% |
| Cells, ***Children*** | 0.4% |
| Animals, ***Biology***, Universe | 0.3% |
| Insects, Cancers | 0.2% |

***Bold**: terms relatively common in abstracts compared to full texts; ***Bold italic***: terms rare in abstracts compared to full texts. These indicate a ranking change of at least 20 in Table 2 compared to Table 3.

## Discussion

The results are limited by the choice of articles and LLM-associated terms. The articles are from a single publisher, and other publishers may influence the language of their authors differently. Individual editorial teams may also influence language use in an individual journal. Moreover, MDPI journals tend to be multidisciplinary and stronger field differences may have been found in a comparison of narrower journals. The strong assumption for some calculations that all language changes from 2021 to 2025 result from LLM is clearly not true since they can also reflect demographic shifts in authorship or the emergence of new topics.

The increased prevalence of most terms assessed in abstracts broadly aligns with previous work (Kousha & Thelwall, 2026), showing that the trends broadly continued until the end of 2024, with decreases occurring for many terms in 2025 compared to 2024, which did not happen for 2024 compared to 2023. The decreases may be due to authors being more careful to avoid high profile LLM-associated terms, different LLMs being used, some of which don't favour a term, or an existing LLM family having its training data or algorithm adjusted in a way that deliberately or accidentally reduces the chance that the term is used. For example, some terms, such as *underscore/s/d/ing*, continued to increase in 2025, whereas some other terms such as *delve/s/d/ing* as widely recognised AI-associated, increased in 2024 and then declined in 2025 (Table 1). Hence, some changes in term use may reflect broader changes in academic writing rather than direct LLM text generation. For instance, some authors, especially from non-English-speaking countries, may use these terms simply because they have learned them through repeated LLM use or because they are seeing them more often in academic texts (e.g., underscore) and consider them acceptable over other similar terms. The results also diverge from previous work (Kousha & Thelwall, 2026) by finding higher proportional increases in LLM term prevalence, perhaps due to the more uniform dataset here.

For field variations, a previous study found relatively low increases 2022-24 in the social sciences for LLM-associated terms in abstracts (Kousha & Thelwall, 2026), whereas the current study found relatively high increases for abstracts for the journal Education (Table 3). This might be a field or publisher difference. The relatively low increase in LLM-associated terms for the life sciences had not been previously noted (Kousha & Thelwall, 2026). *Universe*, which focuses on "theoretical, experimental, and observational progress in fundamental and applied physics", is an outlier for low uptake in full text (Table 2), perhaps due to dense jargon, symbolic reasoning, and mathematical formulations that may be difficult to translate. This area may also be reliant on precise definitions and descriptions, whereas LLMs can tend to vagueness and generality. In addition, theoretical physics papers often seem to avoid standard structures. The one humanities journal that was large enough to include, *Religions*, had a below average increase in LLM term usage (5%). This might be because authorial voice and subtlety of phrasing and argument are valued in the humanities (Hyland, 2004).

The results also (accidentally) suggest that some terms are disappearing because LLMs rarely use them. These may tend to be more archaic terms that are rare outside of academic text, so LLMs may learn them less well than equivalent modern terms that they also learn in other contexts. Perhaps one larger scale impact of LLMs will be to push academic English into more standard English by disfavouring more specialist terms, at least when not essential jargon. Of course, some scholars may use LLMs as a co-author rather than as a translator, creating the potential that reviewers consider legitimate LLM use, perhaps as suggested by the presence of LLM-associated terms, as evidence of misconduct. This is why there have been some automatic attempts to detect LLM authorship (e.g., Xu et al., 2026). The probability estimates here, reported for the first time, show that a reviewer identifying an LLM-associated term almost always has a reasonable (>5%) chance of being wrong if they assume that the term is evidence of LLM use. Even for the cases where the chance of being wrong is under 5% in 2025, it is possible that the probability may increase in the future if, as has happened for other terms, its frequency decreases.

## Conclusions

The results suggest that there are substantial differences between fields in the use of LLMs for writing journal article full texts. The cause of the difference is not clear but may be due to mathematical components that are difficult to translate. If the lack of use in some fields is due to LLM ineffectiveness for the topic then there may be space for LLM developers to fine tune customised versions for these fields.

The results also suggest that analysing LLM term use in abstracts is a reasonably accurate proxy for LLM use in full text overall but not for individual words, and the latter may be ten times greater. Finally, despite the discovery of a few suddenly very popular LLM-associated terms, for most terms it is unsafe to infer LLM use from the appearance of a single LLM-associated term in a text. Some LLM-associated terms may also gradually become part of normal academic writing as authors see them through repeated LLM use or in published papers and then adopt them in their own writing while other terms may become less acceptable over time.

## Declarations

The authors declare no conflict of interest.
ChatGPT was used to write some of the Python programs used for data processing and the first author checked the code and results.

## Appendix

Table A1. The prevalence of the 80 proposed LLM-associated terms in MDPI journal **abstracts** 2021-2025. p(LLM|term) is the probability of LLM influence being the cause of an individual term. The order is the same as for Table 1 (n=237,215 in 2021 to n=264,835 in 2025).

| Proportion | 2021 | 2022 | 2023 | 2024 | 2025 | Increase | Increase ratio | p(LLM|term) |
|---|---|---|---|---|---|---|---|---|
| enhance | 3.1% | 3.2% | 4.9% | 8.9% | 11.6% | 8.6% | 3.8 | 0.74 |
| enhanced | 3.7% | 3.9% | 4.4% | 6.0% | 8.7% | 5.0% | 2.3 | 0.57 |
| comprehensive | 2.9% | 3.1% | 4.8% | 7.8% | 8.3% | 5.4% | 2.9 | 0.65 |
| crucial | 3.3% | 3.5% | 5.8% | 9.9% | 7.9% | 4.6% | 2.4 | 0.58 |
| enhancing | 1.2% | 1.4% | 2.5% | 6.2% | 7.7% | 6.4% | 6.2 | 0.84 |
| robust | 1.7% | 1.7% | 2.0% | 3.0% | 5.1% | 3.4% | 2.9 | 0.66 |
| enhances | 0.6% | 0.7% | 1.1% | 2.7% | 4.3% | 3.7% | 7.1 | 0.86 |
| innovative | 1.5% | 1.5% | 2.1% | 3.3% | 3.2% | 1.7% | 2.2 | 0.54 |
| thus | 4.8% | 4.6% | 4.5% | 3.9% | 2.6% | -2.2% | 0.5 | -0.83 |
| underscore | 0.1% | 0.1% | 0.4% | 1.6% | 2.5% | 2.4% | 31.5 | 0.97 |
| emphasizing | 0.2% | 0.2% | 0.5% | 1.8% | 2.2% | 2.0% | 10.5 | 0.91 |
| underscores | 0.1% | 0.1% | 0.4% | 1.9% | 1.9% | 1.9% | 32.8 | 0.97 |
| facilitate | 1.5% | 1.4% | 1.6% | 1.9% | 1.8% | 0.3% | 1.2 | 0.15 |
| pivotal | 0.5% | 0.4% | 0.9% | 2.0% | 1.7% | 1.2% | 3.3 | 0.70 |
| leveraging | 0.1% | 0.1% | 0.4% | 1.1% | 1.5% | 1.3% | 12.7 | 0.92 |
| facilitating | 0.4% | 0.4% | 0.7% | 1.5% | 1.4% | 0.9% | 3.1 | 0.67 |
| emphasize | 0.4% | 0.4% | 0.5% | 0.9% | 1.3% | 1.0% | 3.6 | 0.72 |
| underscoring | 0.0% | 0.0% | 0.2% | 0.8% | 1.3% | 1.3% | 52.2 | 0.98 |

| word | | | | | | | | |
|---|---|---|---|---|---|---|---|---|
| emphasizes | 0.2% | 0.3% | 0.5% | 1.1% | 1.2% | 0.9% | 5.0 | 0.80 |
| interplay | 0.4% | 0.3% | 0.5% | 1.0% | 1.1% | 0.7% | 2.6 | 0.61 |
| fostering | 0.1% | 0.1% | 0.3% | 0.8% | 1.0% | 0.9% | 8.7 | 0.88 |
| notably | 0.3% | 0.3% | 0.5% | 1.1% | 1.0% | 0.7% | 3.1 | 0.68 |
| facilitates | 0.4% | 0.4% | 0.5% | 0.9% | 1.0% | 0.5% | 2.4 | 0.58 |
| paradigm | 0.6% | 0.6% | 0.5% | 0.6% | 0.9% | 0.3% | 1.6 | 0.37 |
| leverages | 0.1% | 0.1% | 0.2% | 0.6% | 0.9% | 0.8% | 16.0 | 0.94 |
| transformative | 0.1% | 0.1% | 0.2% | 0.4% | 0.8% | 0.7% | 10.0 | 0.90 |
| foster | 0.2% | 0.2% | 0.3% | 0.4% | 0.8% | 0.5% | 3.1 | 0.67 |
| facilitated | 0.4% | 0.3% | 0.5% | 0.7% | 0.7% | 0.3% | 1.8 | 0.45 |
| intricate | 0.1% | 0.1% | 0.6% | 1.3% | 0.6% | 0.5% | 5.2 | 0.81 |
| multifaceted | 0.2% | 0.1% | 0.3% | 0.6% | 0.5% | 0.4% | 3.1 | 0.67 |
| synergy | 0.2% | 0.2% | 0.2% | 0.3% | 0.5% | 0.3% | 2.5 | 0.60 |
| leverage | 0.2% | 0.2% | 0.3% | 0.4% | 0.5% | 0.3% | 3.1 | 0.68 |
| heightened | 0.1% | 0.1% | 0.2% | 0.6% | 0.5% | 0.4% | 6.1 | 0.84 |
| profound | 0.3% | 0.3% | 0.4% | 0.5% | 0.5% | 0.2% | 1.6 | 0.36 |
| emphasized | 0.3% | 0.3% | 0.3% | 0.4% | 0.4% | 0.2% | 1.6 | 0.38 |
| nuanced | 0.1% | 0.0% | 0.1% | 0.4% | 0.3% | 0.3% | 5.9 | 0.83 |
| paradigms | 0.1% | 0.1% | 0.1% | 0.2% | 0.3% | 0.2% | 2.1 | 0.53 |
| delves | 0.0% | 0.0% | 0.3% | 0.8% | 0.2% | 0.2% | 15.9 | 0.94 |
| importantly | 0.4% | 0.4% | 0.3% | 0.3% | 0.2% | -0.2% | 0.6 | -0.75 |
| showcasing | 0.0% | 0.0% | 0.1% | 0.4% | 0.2% | 0.2% | 10.8 | 0.91 |
| navigate | 0.1% | 0.1% | 0.1% | 0.2% | 0.2% | 0.1% | 2.8 | 0.64 |
| fosters | 0.0% | 0.0% | 0.1% | 0.1% | 0.2% | 0.2% | 5.9 | 0.83 |
| navigating | 0.0% | 0.0% | 0.1% | 0.2% | 0.2% | 0.1% | 3.8 | 0.74 |
| leveraged | 0.1% | 0.1% | 0.1% | 0.2% | 0.2% | 0.1% | 1.9 | 0.46 |
| synergies | 0.1% | 0.1% | 0.1% | 0.1% | 0.2% | 0.1% | 1.7 | 0.42 |
| harness | 0.1% | 0.1% | 0.1% | 0.2% | 0.1% | 0.1% | 2.4 | 0.58 |
| underscored | 0.0% | 0.0% | 0.1% | 0.2% | 0.1% | 0.1% | 9.4 | 0.89 |
| harnessing | 0.1% | 0.1% | 0.1% | 0.2% | 0.1% | 0.0% | 1.8 | 0.44 |
| meticulously | 0.0% | 0.0% | 0.1% | 0.3% | 0.1% | 0.1% | 10.3 | 0.90 |
| robustly | 0.1% | 0.1% | 0.1% | 0.1% | 0.1% | 0.0% | 1.3 | 0.24 |
| groundbreaking | 0.0% | 0.0% | 0.1% | 0.1% | 0.1% | 0.1% | 4.9 | 0.80 |
| showcase | 0.1% | 0.1% | 0.1% | 0.2% | 0.1% | 0.0% | 1.4 | 0.29 |
| meticulous | 0.0% | 0.0% | 0.1% | 0.2% | 0.1% | 0.0% | 2.7 | 0.63 |
| bolster | 0.0% | 0.0% | 0.1% | 0.1% | 0.1% | 0.1% | 3.5 | 0.71 |
| delve | 0.0% | 0.0% | 0.2% | 0.3% | 0.1% | 0.0% | 2.7 | 0.64 |
| showcases | 0.0% | 0.0% | 0.1% | 0.1% | 0.1% | 0.0% | 2.9 | 0.66 |
| fostered | 0.0% | 0.0% | 0.0% | 0.0% | 0.1% | 0.0% | 1.5 | 0.35 |
| moreover | 0.2% | 0.2% | 0.2% | 0.1% | 0.1% | -0.1% | 0.4 | -1.45 |
| unveiled | 0.1% | 0.1% | 0.1% | 0.2% | 0.1% | 0.0% | 0.8 | -0.26 |
| furthermore | 0.1% | 0.1% | 0.1% | 0.1% | 0.1% | -0.1% | 0.4 | -1.35 |
| nuances | 0.0% | 0.0% | 0.0% | 0.1% | 0.1% | 0.0% | 2.9 | 0.65 |
| unveil | 0.1% | 0.1% | 0.1% | 0.1% | 0.0% | 0.0% | 0.6 | -0.68 |
| harnessed | 0.0% | 0.0% | 0.1% | 0.1% | 0.0% | 0.0% | 0.9 | -0.10 |
| showcased | 0.0% | 0.0% | 0.1% | 0.1% | 0.0% | 0.0% | 2.1 | 0.51 |
| bolstering | 0.0% | 0.0% | 0.0% | 0.1% | 0.0% | 0.0% | 7.0 | 0.86 |
| heighten | 0.0% | 0.0% | 0.0% | 0.0% | 0.0% | 0.0% | 3.3 | 0.70 |
| unveils | 0.0% | 0.0% | 0.0% | 0.1% | 0.0% | 0.0% | 1.7 | 0.40 |
| paradigmatic | 0.0% | 0.0% | 0.0% | 0.0% | 0.0% | 0.0% | 1.1 | 0.12 |
| crucially | 0.0% | 0.0% | 0.0% | 0.0% | 0.0% | 0.0% | 0.7 | -0.49 |
| delving | 0.0% | 0.0% | 0.0% | 0.1% | 0.0% | 0.0% | 3.3 | 0.70 |
| commendable | 0.0% | 0.0% | 0.0% | 0.1% | 0.0% | 0.0% | 4.4 | 0.77 |
| heightens | 0.0% | 0.0% | 0.0% | 0.0% | 0.0% | 0.0% | 3.8 | 0.74 |
| navigated | 0.0% | 0.0% | 0.0% | 0.0% | 0.0% | 0.0% | 1.6 | 0.36 |
| delved | 0.0% | 0.0% | 0.0% | 0.1% | 0.0% | 0.0% | 10.4 | 0.90 |
| unveiling | 0.0% | 0.0% | 0.0% | 0.1% | 0.0% | 0.0% | 0.7 | -0.35 |
| heightening | 0.0% | 0.0% | 0.0% | 0.0% | 0.0% | 0.0% | 4.6 | 0.78 |
| nuance | 0.0% | 0.0% | 0.0% | 0.0% | 0.0% | 0.0% | 1.8 | 0.45 |
| bolsters | 0.0% | 0.0% | 0.0% | 0.0% | 0.0% | 0.0% | 5.9 | 0.83 |
| navigates | 0.0% | 0.0% | 0.0% | 0.0% | 0.0% | 0.0% | 4.3 | 0.77 |
| complex tapestry | 0.0% | 0.0% | 0.0% | 0.0% | 0.0% | 0.0% | - | 1.00 |
| **[Decr. freq. terms]** | | **50** | **8** | **7** | **45** | | | |